\documentclass{PoS}

\usepackage{amsmath,multirow,arydshln}

\title{Theoretical motivations for precision measurements of oscillation parameters}

\ShortTitle{Theoretical motivations for precision measurements of oscillation parameters}

\author{Silvia Pascoli\\
        IPPP, Department of Physics, Durham University, Durham DH1 3LE, United Kingdom\\
        E-mail: \email{silvia.pascoli@durham.ac.uk}}
        
\author{\speaker{Ye-Ling Zhou}\\
        IPPP, Department of Physics, Durham University, Durham DH1 3LE, United Kingdom\\
        E-mail: \email{ye-ling.zhou@durham.ac.uk}}

\abstract{We discuss two theoretical motivations of precision measurements of oscillation parameters. One is the guidance for flavour model building with flavour symmetries, and the other is the connection with the origin of matter-antimatter asymmetry. We also present our recent progress in these aspects, a new approach of flavour model building based on cross coupling in the flavon potential and a novel mechanism of leptogenesis via phase transition. } 

\FullConference{38th International Conference on High Energy Physics\\
		3-10 August 2016\\
		Chicago, USA}

\begin{document}

\section{Introduction}

Since the discovery of neutrino oscillations, leptonic flavour mixing has been measured to a high level of precision by a series of neutrino oscillation experiments \cite{PDG}. The two mixing angles $\theta_{13}$ and $\theta_{12}$ have been determined precisely by solar and reactor experiments to be $\theta_{13}=8.46^\circ \pm 0.15^\circ$ and $\theta_{12}=33.56^\circ{}^{+0.77^\circ}_{-0.75^\circ}$ in the $1\sigma$ range. The least well known angle is $\theta_{23}$ which is currently measured to be $\theta_{23} = 41.6^\circ{}^{+1.5^\circ}_{-1.2^\circ}$ or $50.0^\circ{}^{+1.1^\circ}_{-1.4^\circ}$ in $1\sigma$, and $(38.4^\circ, 53.0^\circ)$ in $3\sigma$, with the octant - whether $\theta_{23} > 45^\circ$ or $< 45^\circ$, not established. There is a preliminary hint of maximal CP violation with the CP-violating phase $\delta$ around $−90^\circ$, but no constraint in the $3\sigma$ range \cite{globalfit}.

These results strongly suggest specific flavour mixing structures in the lepton sector. Discrete flavour symmetries have emerged as a powerful tool to explain these mixing patterns. In these models the values of the parameters can be predicted or specific correlations between parameters are obtained and the CP-violating phase can take special values, for instance, conserving or maximally-violating values \cite{review}. The precision measurement of mixing angles and the CP-violating phase $\delta$ will be a crucial task in the next-generation neutrino oscillation experiments, such as DUNE and T2HK. They will allow to test the flavour models both by constraining the individual parameters and in probing the correlations among them. 

Hunting for a theory which explains the origin of neutrino masses and mixing, the low energy parameters provide guidance as they are correlated with the ones of the full Lagrangian. In particular, low energy leptonic CP violation is a crucial ingredient as it may share the same origin as that for new particles at high energy \cite{Feruglio:2012cw}, which in many models plays a key role in the generation of the matter-antimatter asymmetry in the early Universe \cite{Fukugita:1986hr}.

Here we discuss the theoretical motivations for the precision measurement of mixing parameters in neutrino oscillations, discussing its impact in flavour model building and the correlation with baryogenesis. We will emphasise a novel mechanism of baryogenesis resulting from a flavon phase transition, in which a strong connection between the lepton asymmetry and flavour models arises.

\section{Flavour models}

\begin{figure}[t]
\centering
\includegraphics[width=0.9\textwidth]{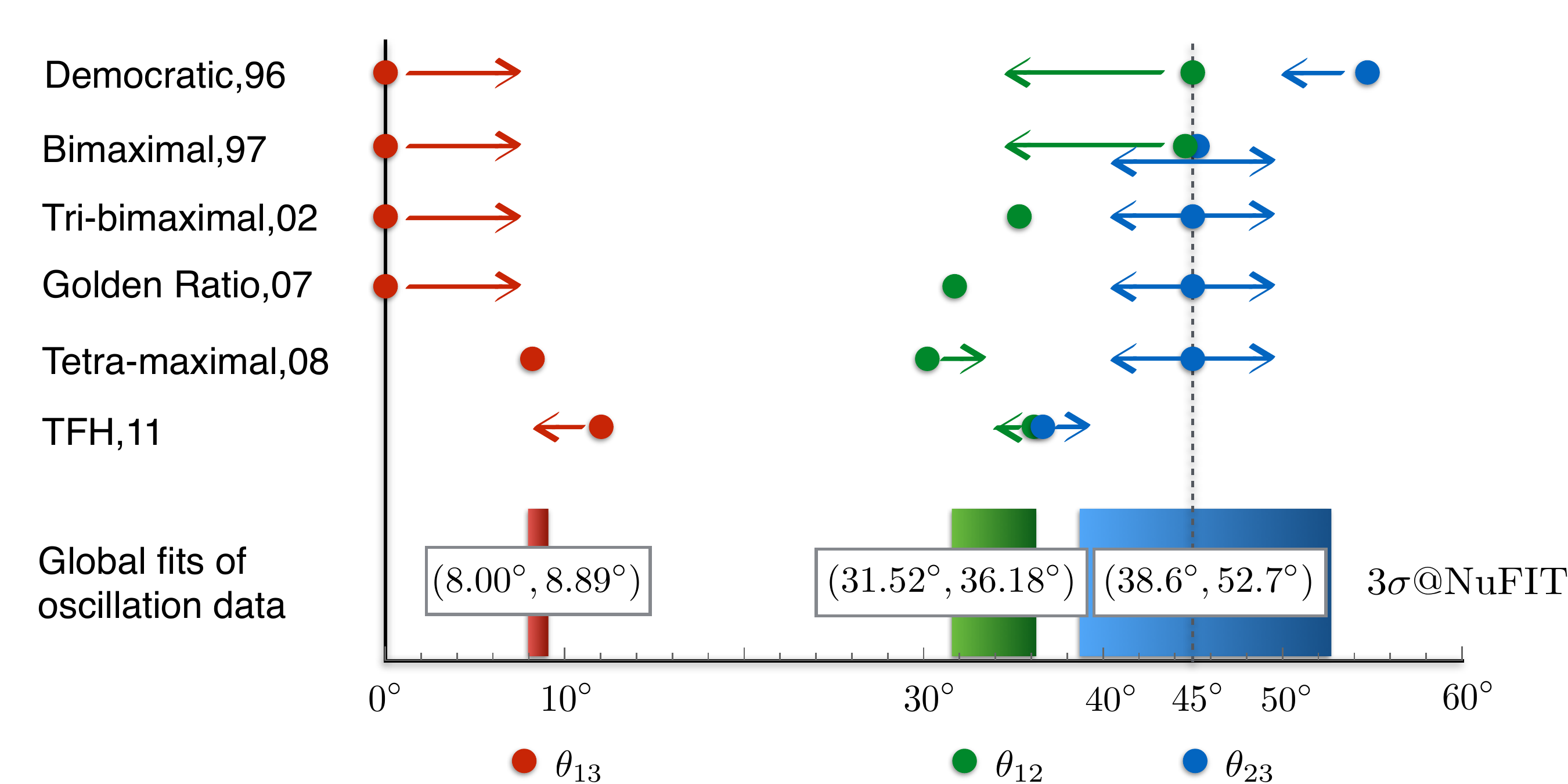}
\caption{\label{fig:fig1} Mixing angles from constant mixing patterns vs oscillation experimental data. }
\end{figure}

Several constant mixing patterns have been proposed, which invoke specific values of the angles, including 
\begin{itemize}
\item democratic mixing \cite{DC}: $\theta_{12} = 45^\circ$ and $\theta_{23} = 54.7^\circ$; 
\item bimaximal mixing \cite{BM}:  $\theta_{12}$ and $\theta_{23}$ are $45^\circ$; 
\item tri-bimaximal (TBM) mixing \cite{TBM}: $\theta_{23} = 45^\circ$ and $\theta_{12} = 35.3^\circ$, very close to current oscillation data; 
\item others as shown in Fig.~\ref{fig:fig1}. 
\end{itemize}
Many of these patterns predict a vanishing $\theta_{13}$, and are therefore excluded by current data \cite{globalfit} or require special large corrections \cite{Xing:2012ej}.

A common way to understand these flavour mixing patterns is to assume an underlying symmetry in the flavour space. In most models, this flavour symmetry is often accompanied by the introduction of some new scalars called flavons. The flavons have specific couplings with leptons satisfying the flavour symmetry. They gain vacuum expectation values (VEVs), leading to the breaking of the flavour symmetry. The flavour mixing is a result of the preserved coupling structures after the breaking. 

We do not know the flavour symmetry. 
It may be Abelian or non-Abelian, continuous or discrete. An Abelian $U(1)$ symmetry may be helpful for realising hierarchical mass structures, such as in the Froggatt-Neilsen mechanism \cite{Froggatt:1978nt}. The discrete case, $Z_n$ symmetries, can be used to realise texture zeros of the mass matrices, which further predict relations between mixing angles and mass ratios \cite{Grimus:2004hf}. A continuous non-Abelian symmetry may predict maximal atmospheric mixing \cite{Alonso:2013nca}. Besides them, non-Abelian discrete symmetries are particularly interesting because they can predict most of the constant mixing patterns in Fig. \ref{fig:fig1} \cite{review}. 

A typical example is the realisation of TBM in the framework of the tetrahedral group $A_4$, with the generators $S$ and $T$ satisfying $S^2=T^3=(ST)^3=1$ \cite{Ma:2001dn}. It has four irreducible representations $\mathbf{1}$, $\mathbf{1'}$, $\mathbf{1}''$ and $\mathbf{3}$. The products of two 3d representations are reduced as $\mathbf{3}\times\mathbf{3}=\mathbf{1}+\mathbf{1}'+\mathbf{1}''+\mathbf{3}_S+\mathbf{3}_A$, where $_S$ and $_A$ denote the symmetric and anti-symmetric parts, respectively. In general, the SM lepton doublets $\ell=(\ell_e, \ell_\mu,\ell_\tau)^T$ are assumed to form a triplet $\mathbf{3}$ of $A_4$. The right-handed (RH) charged leptons $e_R$, $\mu_R$ and $\tau_R$ transform as singlets $\mathbf{1}$, $\mathbf{1}'$ and $\mathbf{1}''$, respectively, and the Higgs $H\sim\mathbf{1}$. We introduce three gauge-invariant flavons $\varphi\sim \mathbf{3}$, $\chi\sim \mathbf{3}$ and $\eta \sim\mathbf{1}$.  
In the type-I seesaw with RH neutrinos $N\sim\mathbf{3}$, the Lagrangian terms for generating lepton masses are given by \cite{A4models}
\begin{eqnarray}
-\mathcal{L}_l &=& \frac{y_e}{\Lambda} (\overline{\ell_L} \varphi)_\mathbf{1} e_R H + \frac{y_\mu}{\Lambda} (\overline{\ell_L} \varphi)_{\mathbf{1}''} \mu_R H + \frac{y_\tau}{\Lambda} (\overline{\ell_L} \varphi)_{\mathbf{1}'} \tau_R H + \text{h.c.} + \cdots \,, \nonumber\\
-\mathcal{L}_\nu &=& y_D (\overline{\ell_L}N)_\mathbf{1} \tilde{H}  + \frac{y_1}{2} \big((\overline{N^c} N)_{\mathbf{3}_S} \chi \big)_\mathbf{1} + \frac{y_2}{2} (\overline{N^c} N)_\mathbf{1} \eta + \text{h.c.} + \cdots \,,
\label{eq:Lagrangian}
\end{eqnarray}
where the dots stand for higher dimensional operators. Once the flavons and the Higgs gain VEVs,
\begin{eqnarray}
\langle \varphi \rangle = (1, 0, 0)^T v_{\varphi} \,,\qquad
\langle \chi \rangle = (1, 1, 1)^T \frac{v_{\chi}}{\sqrt{3}} \,,\qquad
\langle \eta \rangle = v_\eta\,,\qquad
\langle H \rangle = \frac{v_H}{\sqrt{2}}\,,
\label{eq:vevs1}
\end{eqnarray} 
leptons obtain masses and the TBM mixing is realised. 

There are a series of interesting discrete groups, $S_4$, $A_5$, the $\Delta(3n^2)$ and $\Delta(6n^2)$ series, {\it et. al.} \cite{review}. With the assumption of some particular flavon vacuum alignments, most mixing patterns shown in Fig.~\ref{fig:fig1} can be realised. 

One essential question of model building is how to realise the vacuum alignments, i.e., $\langle \varphi \rangle$ and $\langle \chi \rangle$ in Eq.~\eqref{eq:vevs1}.  Different mechanisms have been proposed to solve it \cite{A4models}, but inevitably lots of new degrees of freedom have to be introduced, which are not essential for the realisation of flavour mixing. 
In Ref.~\cite{Pascoli:2016eld}, we find that such simple vacuum alignments may be a result of the spontaneous breaking of the flavour symmetry. For example, in $A_4$ models, the VEVs of $\varphi$ and $\chi$ in Eq.~\eqref{eq:vevs1} can be directly obtained from their self couplings in the potential. The cross coupling between $\varphi$ and $\chi$, which modifies the vacuum alignments, cannot be forbidden. Fortunately, once we assume it to be small due to phenomenological considerations, it shifts the VEVs slightly and gives rise to $\theta_{13}$ and CP phase $\delta$. In the simplest case ($\varphi_1=\varphi_1^*$, $\varphi_2=\varphi_3^*$, $\chi_1=\chi_1^*$ and $\chi_2=\chi_3^*$), the VEVs are shifted to
\begin{eqnarray}
\langle \varphi \rangle \approx (1, \epsilon_{\varphi}, \epsilon_{\varphi}^*)^T v_\varphi \,,\quad
\langle \chi \rangle \approx (1-2 \epsilon_\chi, 1+\epsilon_\chi, 1+\epsilon_\chi)^T \frac{v_\chi}{\sqrt{3}} \,, 
\label{eq:flavon_vevs}
\end{eqnarray}
and non-zero $\theta_{13}$, $\delta$ are obtained with expressions 
$\sin\theta_{13} \approx |\text{Re}{\epsilon_\varphi}|$ and
$\delta \approx \mp 90^\circ -2 \text{Re}\epsilon_\varphi$, 
where $\epsilon_\varphi $ is complex and $\epsilon_\chi$ is real. 
Almost maximal CP violation is predicted, and a new sum rule
\begin{eqnarray}
\delta \approx \mp( 90^\circ + \sqrt{2} \theta_{13})\,. 
\label{eq:sumrule_ra}
\end{eqnarray}  
is found \cite{Pascoli:2016eld}. The mixing angles $\theta_{12}$ and $\theta_{23}$ also gain small corrections from their leading order values. This model can be tested in the future oscillation experiments. 

In the framework of flavour symmetries, the CP symmetry can be extended to the generalised CP (GCP). The basic idea is that for a multiplet $\varphi$ in the flavour space, the CP transformation should be a combination of the traditional CP and a basis transformation in the flavour space, i.e.,
\begin{eqnarray}
\text{GCP: } \varphi \to X \varphi^* \,.
\end{eqnarray} 
Here, $X$ is called the GCP transformation matrix. It is a unitary matrix, not arbitrary, but must satisfy the consistency condition $\rho(g')=X\rho(g)X^{-1}$, where $\rho(g')$ and $\rho(g)$ are representation matrices of the group elements $g'$ and $g$, respectively \cite{Holthausen:2012dk}. Due to this condition, once the flavour symmetry is fixed, all the candidates of GCP transformations will be determined. This approach makes powerful predictions, with only one free parameter in the mixing matrix \cite{Feruglio:2012cw}. 
A theory with GCP symmetries does not mean that it preserves the ordinary CP symmetry unless $X = \mathbf{1}$. One famous example is the so-called $\mu$-$\tau$ reflection symmetry \cite{mutau}, which corresponds to the transformation $e\leftrightarrow e^c$, $\mu\leftrightarrow \tau^c$. It gives rise to the maximal atmospheric mixing and maximal CP violation $\delta=\pm90^\circ$. This symmetry can be predicted by $A_4$, or any other groups containing $A_4$ \cite{review}.

\section{Lepton asymmetry}

Precision measurements of mixing angles and the CP phase $\delta$ may help us to understand the origin of the matter-antimatter asymmetry in the Universe. The observed ratio of the baryon to photon is $\eta_B\approx 6.2\times10^{-10}$ \cite{PDG}. One of the most studied mechanisms for its generation is leptogenesis. 

In the typical implementation of leptogenesis in type-I seesaw models \cite{Fukugita:1986hr}, the decay of RH neutrinos produces a lepton asymmetry which is later partly converted into a baryon asymmetry through Sphaleron processes. The lepton asymmetry $\Delta f_{\ell \vec{k}} \equiv f_{\ell \vec{k}} - f_{\overline{\ell} \vec{k}}$ directly depends on the asymmetry between the decay rate of $N\to \ell_\alpha H$  and of the CP-conjugate process $N\to \overline{\ell_\alpha} H^*$. Ignoring flavour effects,
\begin{eqnarray}
\Delta f_{\ell \vec{k}} \propto \sum_{i=2,3}\text{Im}\left\{ [(Y^\dag Y)_{1i}]^2 \right\} \frac{M_1}{M_i}\,, 
\end{eqnarray}
where $Y_{\alpha i}$ is the Yukawa coupling coefficient between $\ell_\alpha$ and RH neutrino mass eigenstate $N_i$ with $N_1$ the lightest one. 
To generate a non-zero $\Delta f_{\ell \vec{k}}$, the condition $\text{Im}\{ [(Y^\dag Y)_{1i}]^2 \}\neq 0$ is required. 

In most flavour models based on non-Abelian discrete flavour symmetries, the above condition is not satisfied. The reason is that these models prefer a special feature of $Y$. For example, $Y$ from Eq.~\eqref{eq:Lagrangian} is given by $Y=(y_\text{D} v_H/\sqrt{2} )U_\ell^\dag \mathbf{1}_{3\times 3} U_\nu $, where $U_\ell$ and $U_\nu$ are unitary matrices diagonalising the charged lepton and RH neutrino mass matrices, respectively. 
While these models are more predictive, they imply $\text{Im}\{ [(Y^\dag Y)_{1i}]^2 \}=0$. A lepton asymmetry could still generated at higher orders. 

Recently, we have proposed a new mechanism which shows strong connection between the lepton asymmetry and flavour models \cite{Pascoli:2016gkf}. 
Different from thermal leptogenesis, this mechanism does not generate the lepton asymmetry through the decay of any heavy particles, but from a CP-violating phase transition. In this new mechanism, we do not specify any neutrino mass origins, but require the coefficients $\lambda_{\alpha\beta}$ of the lepton-number-violating Weinberg operator $({\lambda_{\alpha\beta}}/{\Lambda}) \overline{\ell_{\alpha L}} \tilde{H} \tilde{H}^T \ell_{\beta L}^c + \text{h.c.}$
be dynamically realised during the phase transition of a flavon $\phi$, 
\begin{eqnarray}
\lambda_{\alpha\beta} = \lambda_{\alpha\beta}^0 + \lambda_{\alpha\beta}^1 \frac{\langle \phi \rangle}{v_\phi}\,.
\end{eqnarray}
During the vacuum phase transition from the trivial phase $\langle \phi \rangle=0$ to the phase $\langle \phi \rangle=v_\phi$, the coupling $\lambda_{\alpha\beta}$ is time-dependent, the lepton asymmetry is generated via the interference of the Weinberg operator at different times. 
Technically, to calculate this asymmetry, we work in the closed time path formalism, and the lepton asymmetry is obtained from a 2-loop self-energy correction,  
\begin{eqnarray}
\Delta f_{\ell \vec{k}} = \frac{3\,\text{Im}\left\{ \text{tr}[ m^{0*}_{\nu}m^{}_{\nu} ]  \right\}T^2}{\left( 2\pi \right)^{4}v^{4}_{H}}  F\left( \frac{k}{2T},\frac{\gamma}{T} \right)\,.
\end{eqnarray}
where $m_\nu^0={\lambda^0} v_H^2 / {\Lambda}$, $m_\nu={\lambda} v_H^2 / {\Lambda}$, $\gamma$  is the sum of the thermal widths of the Higgs and the leptons, and $F( {k}/{2T},{\gamma}/{T})$ is a loop factor. 
As shown in Ref.~\cite{Pascoli:2016gkf}, the loop factor provides an $\mathcal{O}(10)$ factor enhancement. While $m_\nu$ is identical to the active neutrino mass matrix, $m_\nu^0$ is an effective mass parameter corresponding to the flavour structure before $\phi$ gets a VEV, which is strongly dependent upon the flavour model construction and worth further studies. Without loss of generality, one may assume $\text{Im}\{\text{tr}[m_\nu^{0} m_\nu^*]\}$ to be of the same order as $m_\nu^2 \sim (0.1~\text{eV})^2$. 
Then, the temperature for the phase transition approximates to
$T \sim 10 \sqrt{\Delta f_{\ell \vec{k}}}\, {v_H^2}/{m_\nu} $. 
Requiring $\Delta f_{\ell \vec{k}}\sim\eta_B$, we conclude the temperature for successfully baryogenesis is $T\sim 10^{11}$ GeV. 

\section{Summary}

Precision measurements of the mixing angles and the CP violation can guide flavour model building and may be helpful for understanding the mysteries of the baryon asymmetry in the observed Universe. We have reviewed how the observed mixing pattern can be realised using discrete flavour symmetries. We have focussed on a new approach of flavour model building based on flavon cross couplings, which break some of the residual discrete symmetries and reconcile model predictions with the measured values of the mixing angles, without introducing additional degrees of freedom. A sum rule between $\theta_{13}$ and $\delta$ is obtained and almost maximal CP violation is predicted. In the end, we have presented a novel mechanism of leptogenesis via the CP-violating phase transition. It does not require a specific model for neutrino mass generation. This mechanism shows a strong connection between the lepton asymmetry and flavour models which is worth further studies.

{\bf Acknowledgement. } We are grateful for financial support from the European Research Council under ERC Grant ``NuMass'' (FP7-IDEAS-ERC ERC-CG 617143). SP gratefully acknowledges partial support from the Wolfson Foundation and the Royal Society.


\begin{thebibliography}{99}

\bibitem{PDG}
  K.~A.~Olive {\it et al.} [Particle Data Group Collaboration],
  Chin.\ Phys.\ C {\bf 38}, 090001 (2014).

\bibitem{globalfit}
  I.~Esteban, M.~C.~Gonzalez-Garcia, M.~Maltoni, I.~Martinez-Soler and T.~Schwetz,
  arXiv:1611.01514 [hep-ph].

\bibitem{review} 
  For some reviews, see e.g., 
  G.~Altarelli and F.~Feruglio,
  Rev.\ Mod.\ Phys.\  {\bf 82}, 2701 (2010);
  S.~F.~King, A.~Merle, S.~Morisi, Y.~Shimizu and M.~Tanimoto,
  New J.\ Phys.\  {\bf 16}, 045018 (2014).

\bibitem{Feruglio:2012cw} 
  F.~Feruglio, C.~Hagedorn and R.~Ziegler,
  JHEP {\bf 1307}, 027 (2013).
  
\bibitem{Fukugita:1986hr} 
  M.~Fukugita and T.~Yanagida,
  Phys.\ Lett.\ B {\bf 174}, 45 (1986).
   
\bibitem{DC}
  H.~Fritzsch and Z.~Z.~Xing,
  Phys.\ Lett.\ B {\bf 372}, 265 (1996);
  Phys.\ Lett.\ B {\bf 440}, 313 (1998).
  
\bibitem{BM}
  F.~Vissani,
  hep-ph/9708483;
  V.~D.~Barger, S.~Pakvasa, T.~J.~Weiler and K.~Whisnant,
  Phys.\ Lett.\ B {\bf 437}, 107 (1998).
  
\bibitem{TBM}
  P.~F.~Harrison, D.~H.~Perkins and W.~G.~Scott,
  Phys.\ Lett.\ B {\bf 530}, 167 (2002);
  Z. Z.~Xing,
  Phys.\ Lett.\ B {\bf 533}, 85 (2002);
  P.~F.~Harrison and W.~G.~Scott,
  Phys.\ Lett.\ B {\bf 535}, 163 (2002);
  X.~G.~He and A.~Zee,
  Phys.\ Lett.\ B {\bf 560}, 87 (2003).

\bibitem{Xing:2012ej} 
  Z.~Z.~Xing,
  Chin.\ Phys.\ C {\bf 36}, 281 (2012).

\bibitem{Froggatt:1978nt} 
  C.~D.~Froggatt and H.~B.~Nielsen,
  Nucl.\ Phys.\ B {\bf 147}, 277 (1979).

\bibitem{Grimus:2004hf} 
  W.~Grimus, A.~S.~Joshipura, L.~Lavoura and M.~Tanimoto,
  Eur.\ Phys.\ J.\ C {\bf 36}, 227 (2004).

\bibitem{Alonso:2013nca} 
  R.~Alonso, M.~B.~Gavela, G.~Isidori and L.~Maiani,
  JHEP {\bf 1311}, 187 (2013).

\bibitem{Ma:2001dn} 
  E.~Ma and G.~Rajasekaran,
  Phys.\ Rev.\ D {\bf 64}, 113012 (2001);
 
\bibitem{A4models}
  G.~Altarelli and F.~Feruglio,
  Nucl.\ Phys.\ B {\bf 720}, 64 (2005);
  Nucl.\ Phys.\ B {\bf 741}, 215 (2006).
  
\bibitem{Pascoli:2016eld} 
  S.~Pascoli and Y.~L.~Zhou,
  JHEP {\bf 1606}, 073 (2016).

\bibitem{Holthausen:2012dk} 
  M.~Holthausen, M.~Lindner and M.~A.~Schmidt,
  JHEP {\bf 1304}, 122 (2013).

\bibitem{mutau}
  K.~S.~Babu, E.~Ma and J.~W.~F.~Valle,
  Phys.\ Lett.\ B {\bf 552}, 207 (2003);
  P.~F.~Harrison and W.~G.~Scott,
  Phys.\ Lett.\ B {\bf 547}, 219 (2002).
  
\bibitem{Pascoli:2016gkf} 
  S.~Pascoli, J.~Turner and Y.~L.~Zhou,
  arXiv:1609.07969 [hep-ph].
  
\end{thebibliography}
\end{document}